\documentclass[a4paper,11pt]{article}
\pdfoutput=1

\usepackage{jinstpub}
\usepackage{xspace}
\usepackage{float}

\usepackage[T1]{fontenc}

\newcommand{\pt}{$\mathrm{p_{T}}$\xspace}
\newcommand{\met}{$\mathrm{E_{T}^{miss}}$\xspace}
\newcommand{\ttbar}{$\mathrm{t\bar{t}}$\xspace}
\newcommand{\ttg}{$\mathrm{t\bar{t}\gamma}$\xspace}
\newcommand{\st}{$\mathrm{t(\bar{t})}$\xspace}
\newcommand{\stg}{$\mathrm{t(\bar{t})\gamma}$\xspace}
\newcommand{\dr}{$\mathrm{\Delta R}$\xspace}
\newcommand{\nupz}{$\mathrm{p_{z}^{\nu}}$\xspace}

\title{\boldmath Photon radiation effects in kinematic
reconstruction of top quarks}

\author{D. Dobur,}
\author{J. Knolle,}
\author{G. Mestdach}
\author[1]{and K. Skovpen\note{Corresponding author.}}

\affiliation{Ghent University,\\Sint-Pietersnieuwstraat 33, 9000 Gent}

\emailAdd{kirill.skovpen@cern.ch}

\abstract{Kinematic reconstruction of top quarks allows to
define a set of kinematic observables relevant to various
physics processes that involve top quarks and provides an
additional handle for the suppression of background events.
Radiation of photons in association with the top quarks alters
the kinematics and the topology of the event, leading to 
visible systematic effects in measurable observables. 
The present study introduces an improved reconstruction of the top quark 
kinematics in the presence of photon radiation. The results
are presented for processes with top quark pair production, as
well as for singly-produced top quarks.}

\begin{document} 
\maketitle
\flushbottom

\newpage

\section{Introduction}
\label{sec:intro}

The measurement of the top quark production cross sections, as well as the
study of the properties of top quarks, remain a foreground direction 
in a large number of experimental studies across several
experiments at the Large Hadron Collider (LHC) at CERN. Top quarks are
copiously produced at the LHC, allowing to achieve an unprecedented
precision in the measurement of the inclusive production cross
sections of top quark pairs (\ttbar) down to the
level of $\simeq$~2\%~\cite{TTbarXsecATLAS}. The production cross section
of a single top quark (\st) in the predominant $t$ channel has been measured with
the precision of $\simeq$~15\%~\cite{STXsecCMS}. In addition to the measurement of the
inclusive cross sections, the amount of recorded data at the LHC
allows to perform measurements of the differential
production cross sections as a function of various kinematic observables.

The electroweak couplings of the top quark can be directly probed in processes 
involving the associated production of top quarks with vector bosons.
The production of top quarks in association with a photon (\ttg) provides 
direct access to the study of the electromagnetic couplings of the top
quark at hadron colliders~\cite{BauDir,BauPro,BevOff,BevOff2,BevOff3,PagAut}.
With respect to \ttbar production, the top quark charge asymmetry
is particularly enhanced to the level of $\simeq$~10\% in the
$\mathrm{q\bar{q}}$-initiated production channel of the \ttg
process~\cite{TTGChargeAsym}, which is significantly larger than in
\ttbar. Potential new physics effects can be induced as anomalous electric
dipole moments, modifying the standard model (SM) predictions for
various kinematic observables for this process~\cite{AguMin,BouMin}.
Within the effective field theory (EFT) approach, the potential
modifications of the structure of the \ttg interaction vertex are parametrized 
with the respective dimension-six operators~\cite{LHCtopWGEFT}.

The \ttg process was experimentally studied at CDF~\cite{AalEvi},
ATLAS~\cite{AadObs,AabMea,AabMea2,AabMea3} and
CMS~\cite{SirMea,SirMea2}. The measured cross sections in semileptonic
and dilepton channels show good agreement with the SM predictions.
The differential cross section is measured as a function of multiple kinematic 
variables involving leptons and radiated photons. Several of these
variables, such as the angular separation between the two leptons, which 
can be reconstructed in the dilepton channel, are particularly 
sensitive to \ttbar spin correlation effects. The charge asymmetry has not
yet been studied experimentally, as it requires a good handle to
separate the $\mathrm{gg}$- and $\mathrm{q\bar{q}}$-initiated \ttg production 
processes~\cite{TTGChargeAsym}. 
It was shown that the photon rapidity can potentially allow to
distinguish between these two processes in the measured differential
distributions. The measurement of the \ttbar charge asymmetry in \ttg
events however requires a full reconstruction of the \ttg final state, which has not yet
been established in experimental studies.

In addition to the \ttbar process, the photons can also be
radiated in processes involving the production of single top quarks, \stg~\cite{TG1,TG2}.
This process has been studied by the CMS experiment leading to the observed
evidence of 4.4 standard deviations~\cite{TGCMS}. The
production mechanism of \stg is dominated by the electroweak $t$ channel process,
where the light-flavour quark in the hard-scattering process recoils against the 
top quark to produce an energetic jet in the forward region of the detector. 
As in \ttg production, the \stg process is sensitive to the
electromagnetic dipole moments of the top quark and potential new
physics effects described within an EFT approach.

\section{Photon radiation in processes with top quarks}
\label{sec:kinematics}

The performed study aims at reconstructing top quarks in the 
presence of photon radiation in several event topologies 
including single-top and \ttbar production channels. 
The photon can be emitted from any charged particle that is involved in the
process, from the products of subsequent decays of unstable
particles, as well as in the parton shower process.
The proposed algorithm of the kinematic reconstruction targets at
properly assigning the origin of the emitted photon. The study
primarily focuses on the processes with the production of top quarks, where a photon is
radiated at the matrix-element (ME) level.
The photon can be radiated from the initial state quarks, the
intermediate top quarks and W bosons, as well as from the
final-state quarks and leptons. The kinematic effects 
that are induced by the photon radiation in the
\ttbar process were previously discussed in Ref.~\cite{BevOff},
indicating a significant fraction of events with photon emission
from top quark decays, especially in the low photon \pt
region. The inclusion of radiated photons in the reconstruction of
events with top quarks is therefore crucial in order to properly define the top
quark-related kinematic observables at the reconstruction level in the
measurement of differential cross sections, as well as in studies of the
\ttbar charge asymmetry and spin correlation effects. 

Leptonic decays of W bosons originating from top quark decays 
lead to the presence of a neutrino (or an anti-neutrino) in events,
which can only be partially reconstructed at hadron colliders.
The longitudinal component of the neutrino's momentum (\nupz) remains
unconstrained. However, it can be inferred by imposing a set of kinematic constraints on
the topology of the reconstructed events that describe the top quark
decay kinematics. The \nupz can be derived by resolving
a quadratic equation involving constraints on the
reconstructed masses of the W boson and the top quark. The W
boson mass is reconstructed from a lepton and a neutrino ($\mathrm{W^{-}
\to \ell^{-}\bar{\nu}}$, $\mathrm{W^{+} \to \ell^{+}\nu}$), while the
reconstruction of the top quark mass additionally involves the hadronic jet
that is associated with the decay of a heavy hadron, a b
jet ($\mathrm{t \to \ell^{+}\nu}b$, $\mathrm{\bar{t} \to
\ell^{-}\bar{\nu}}\bar{b}$). The missing transverse energy,
(\met) which is defined as the negative vectorial sum of all
reconstructed objects in the event in the transverse plane of the colliding beams,
provides crucial information about the transverse component of the
neutrino's momentum. In the case of the fully hadronic decays of the top quarks, the
reconstruction of the top quark mass involves three jets, one of which
is identified as a b jet. The dominant contribution to the measured experimental
resolution in the reconstructed spectra of the W boson and the top
quark masses is therefore predominantly described by the jet energy 
scale and resolution effects.
The production of single top quarks leads to the presence of either a leptonic or
a hadronic W boson decay, with an additional b quark arising from the
top quark decay. 

The top quarks that are produced in pairs lead to
two distinctive event topologies. The semileptonic \ttbar events contain one
leptonic decay of a W boson, with the other W boson decaying
hadronically. The dilepton final states arise from leptonic decays
of two W bosons, subsequently producing two neutrinos in the final state. 
The kinematic properties of the described event topologies can
be fully established, if the corresponding leptons, jets and
\met are well reconstructed in the event.

The kinematic reconstruction of top quarks is performed in \ttg semileptonic and
dilepton events, as well as in events with an associative production of 
single top quarks and a photon with leptonic W boson decays.
Events are simulated with the $\mathrm{MadGraph5\_aMC@NLO}$
v2.8.2~\cite{AlwAut} generator at leading order, interfaced with the Pythia
v8.1~\cite{SjoAut} program for the simulation of hadronization and fragmentation
processes. At ME level, the event generation corresponds
to 2 $\to$ 7 and 2 $\to$ 5 production processes for \ttg and \stg,
respectively, where a photon can be radiated from any charged particle
involved in the process. Event generation involves on-shell top
quarks, and it accounts for interference effects in the relevant production diagrams.
The photon origin is determined by looking at its ancestors in the
generated event record.
Angular separation between a photon with a minimum \pt
of 5 GeV and any other final-state particle is required to be \dr =
$\mathrm{\sqrt{(\Delta \phi)^{2} + (\Delta \eta)^{2}}}$ $>$ 0.3. The generated events are 
passed through a simplified detector simulation 
that is realized within the Delphes v.3.4.2~\cite{FavDel} package that
models a realistic detector response and performs the event reconstruction.
Reconstructed events that correspond to at least one top quark decaying
via a leptonically decaying W boson are required to have at least one lepton (electron or
muon) and one photon with \pt $>$ 20 GeV and an absolute pseudorapidity, |$\eta$|, of less
than 2.4. Jets are reconstructed using the anti-$\mathrm{k_{T}}$ jet
algorithm~\cite{Antikt} with R = 0.4 and are selected with \pt $>$ 30
GeV and |$\eta$| $<$ 2.4.
It is additionally required that jets do not have a nearby lepton that
is reconstructed within \dr = 0.4. 
The photon must be separated from any lepton or
jet object by \dr > 0.4. The event selection applied in the dilepton channel requires the
presence of at least two leptons in the final state. The \met is
reconstructed from tracks and particle-flow calorimeter deposits. 
The reconstructed leptons, jets and the photon, are geometrically
matched within \dr = 0.4 to the corresponding generated particles that
are associated with the parton-level production and decay processes. 
Selected events where all generated particles are properly matched to the
corresponding reconstructed objects are used to perform the study of the 
kinematic reconstruction.

\section{Kinematic event reconstruction}
\label{sec:kin}

\begin{figure}[!htbp]
\centering
\includegraphics[width=.49\textwidth]{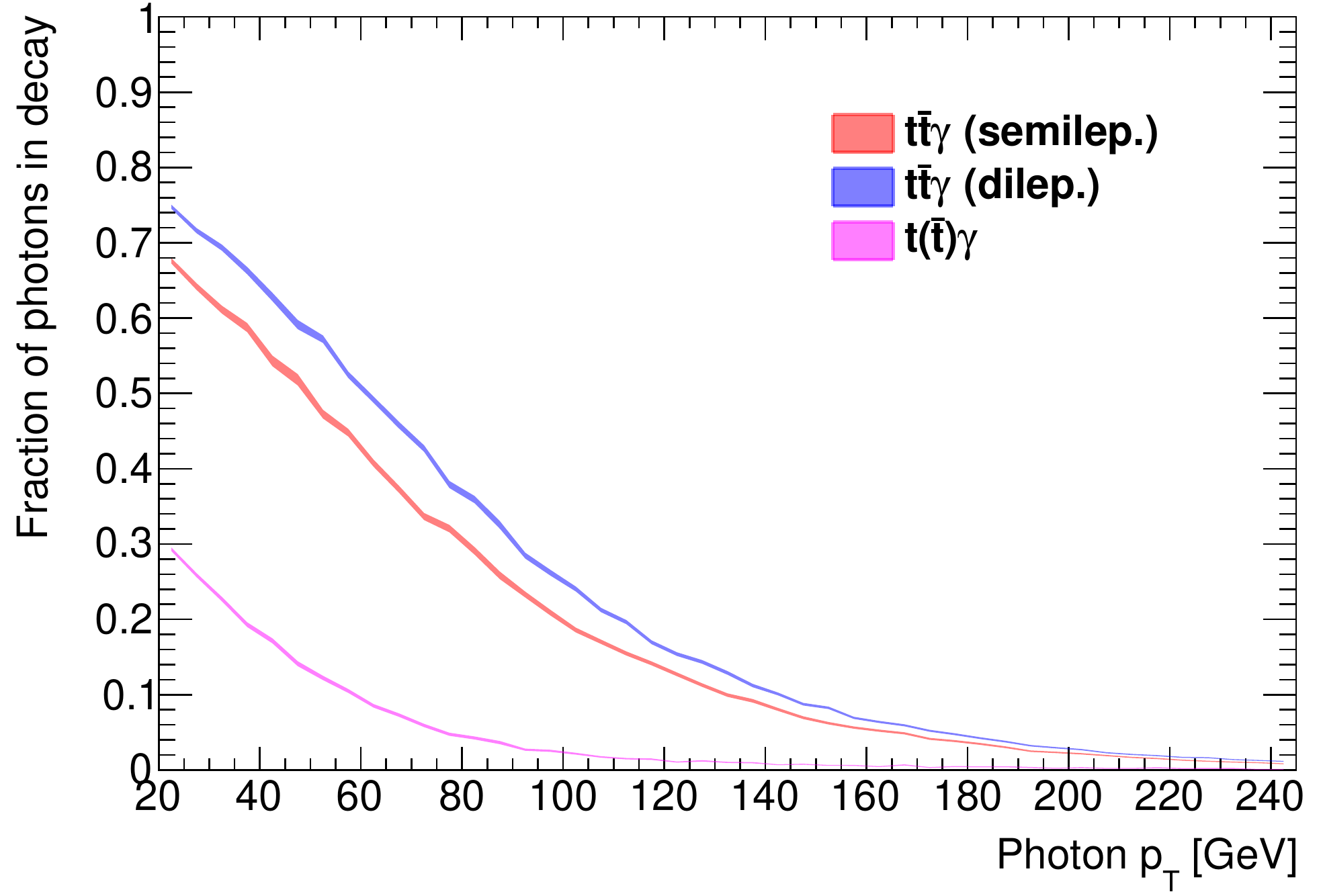}
\hfill
\includegraphics[width=.49\textwidth]{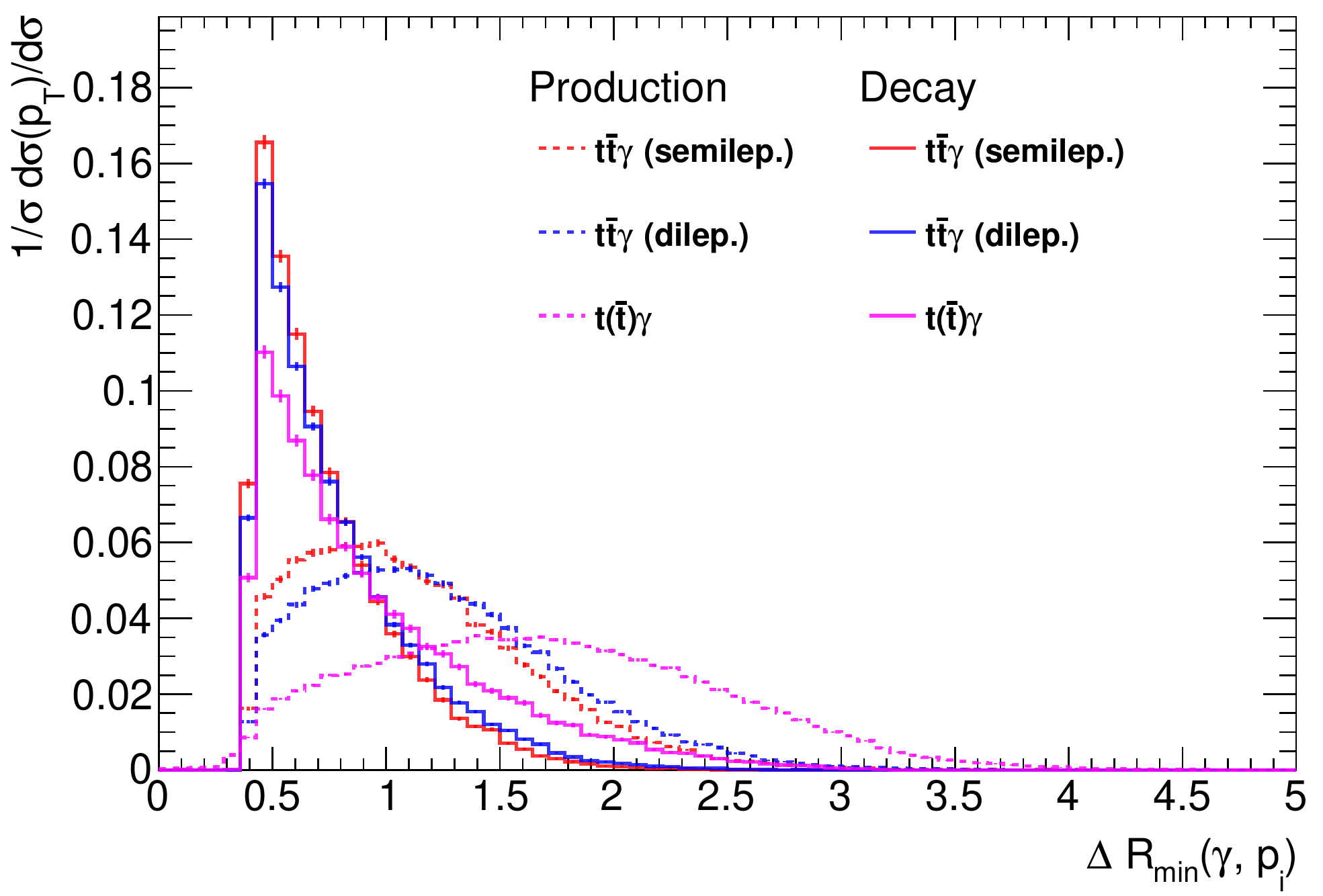}
\caption{\label{fig:photonOrigin} Left: Relative fraction of events with
a photon that is radiated from the decay products of top quarks. Right: Minimal \dr separation between 
the photon and any of the parton-level final state particles.}
\end{figure}

Events that satisfy the described selection criteria are used to perform a kinematic 
reconstruction of the \ttg and \stg processes, and are classified into
event categories where the photon is radiated either from the
production part of the parton-level process (``production''), or
from the decay products of top quarks (``decay''). The number of
reconstructed objects used as input to the kinematic fit corresponds
to the number of generated particles at the ME level. 
The fraction of the events where a photon is emitted in the decay, and the
distributions representing an angular separation between the 
photon and the closest parton-level final state particle,
are presented in Fig.~\ref{fig:photonOrigin}. The radiation of photons with high
momentum is kinematically enhanced in the production case, while the
emitted photons in the top quark decay are mainly radiated at small angles 
and generally have a softer \pt spectrum. The observed difference in
the measured event fractions between the semileptonic and the dilepton
channels is due to the selection criteria applied on the \dr
separation between a reconstructed photon and other final-state objects.

The reconstruction of the neutrino's longitudinal momentum in the context of \ttbar events,
covering both the semileptonic and the dilepton final
states, is discussed in detail in Refs.~\cite{AbbMea, AbuMea, AbaPre, AbaMea, ErdLik, ErdFro, CasRec,
KviStu, SyeBid}, with the analytic solutions for \nupz derived in
Refs.~\cite{BetAna, DemRec, SonAlg}. The methodology of the presented
study does not directly use the proposed analytic solutions, but rather attempts at
resolving the system of the relevant unknowns by performing a likelihood fit
based on the initial set of equations involving \nupz. 
A suitable solution for \nupz is determined in a
log-likelihood fit that is implemented in a dedicated
\texttt{TopPhit} package~\cite{TopPhit}, which makes use of the ROOT
toolkit libraries~\cite{ROOT}. The fit is based on the comparison of
the reconstructed top quark and the W boson masses with the corresponding
truth-level values. The general likelihood function that is used to perform 
a fit is defined as:

\begin{equation}
\mathrm{\mathcal{L} = -2 ln \bigg[\prod_{i} P_{i}(rec|gen) \times P_{neg}\bigg]},
\end{equation}

\noindent where the $\mathrm{P_{i}}$ denote the
probabilities of the reconstructed masses of a top quark and a W boson
to correspond to their predicted values from generator, following the
Breit-Wigner distribution. 

The likelihood function contains four main 
terms that correspond to the top quark, the top
antiquark, and two W bosons in the kinematic reconstruction of 
\ttg events. There is a single neutrino produced in semileptonic
\ttg events, and therefore the \nupz remains the only unknown. As a
consequence of resolving a quadratic equation, there are two possible
\nupz solutions. The ambiguity is resolved by imposing
a constraint on the reconstructed top quark mass value, in addition to
the W boson mass requirement. The hadronic decay of the top quark in
\ttg semileptonic events is reconstructed using the corresponding W boson and
top quark mass requirements, leading to four constraints, together
with the reconstruction of the leptonic top quark decay. There are also
four main likelihood terms included in the kinematic fit of dileptonic \ttg
events. However, in this case there are two neutrinos with four
unknowns in the reconstruction. With four terms corresponding to the W boson 
and top quark masses included in the
likelihood fit, the minimized function has zero degrees of freedom in the
fit. There are two likelihood terms that are included in the \stg case,
similar to the reconstruction of the leptonic decay of the top quark 
in the \ttg semileptonic system.
The last term ($\mathrm{P_{neg}}$) in the likelihood definition refers to 
the additional penalty term against the nonphysical
solutions associated with imaginary values that can appear in the
derivation of the neutrino's longitudinal momentum.
This penalty term is defined as the sum of absolute values of square
roots giving imaginary solutions in event, guiding the minimization
procedure towards the correct minima.

\begin{figure}[!htbp]
\centering
\includegraphics[width=.99\textwidth]{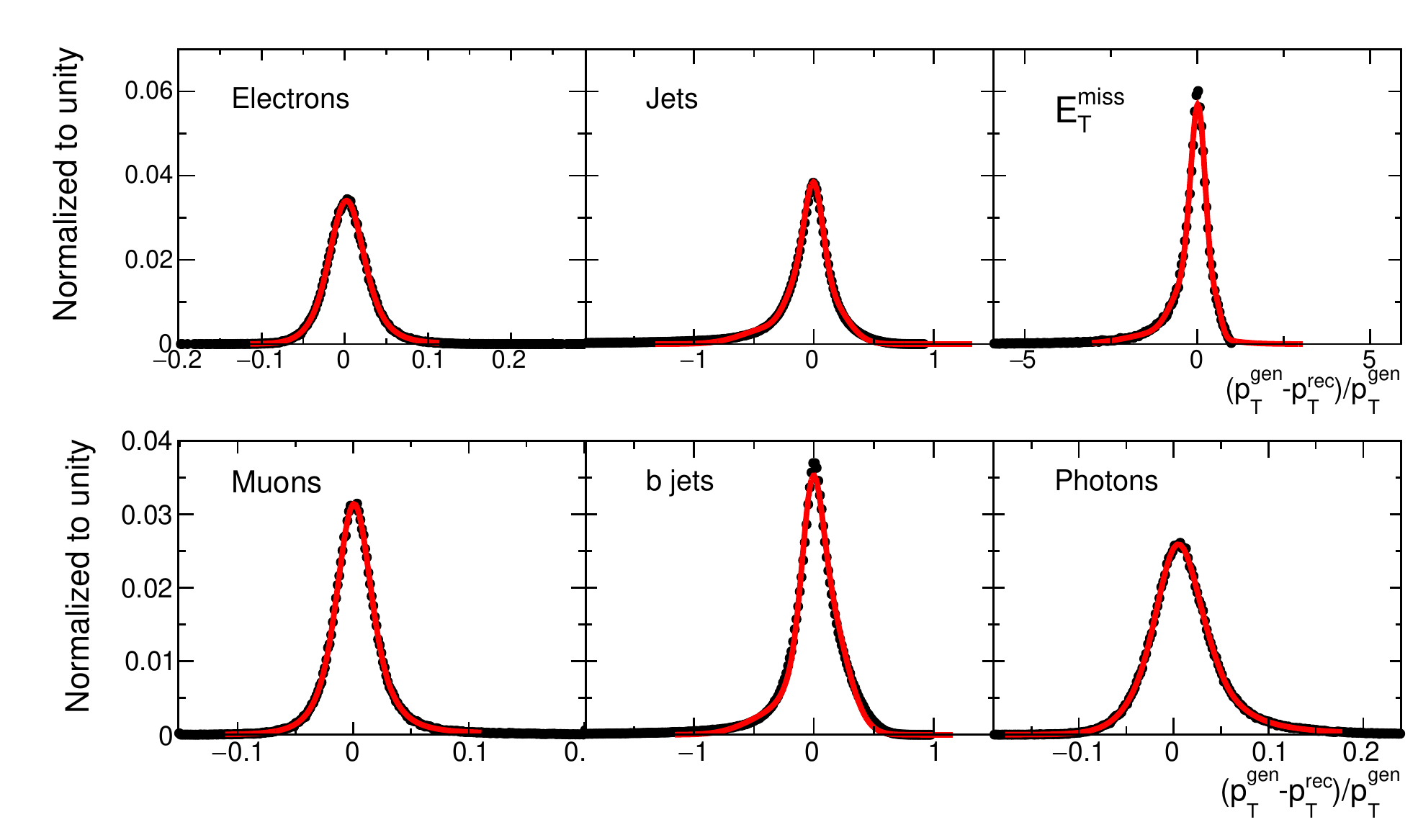}
\caption{\label{fig:tf} Some examples of transfer functions used in
the kinematic reconstruction that are measured for
various reconstructed objects.}
\end{figure}

\begin{figure}[!htbp]
\centering
\includegraphics[width=.80\textwidth]{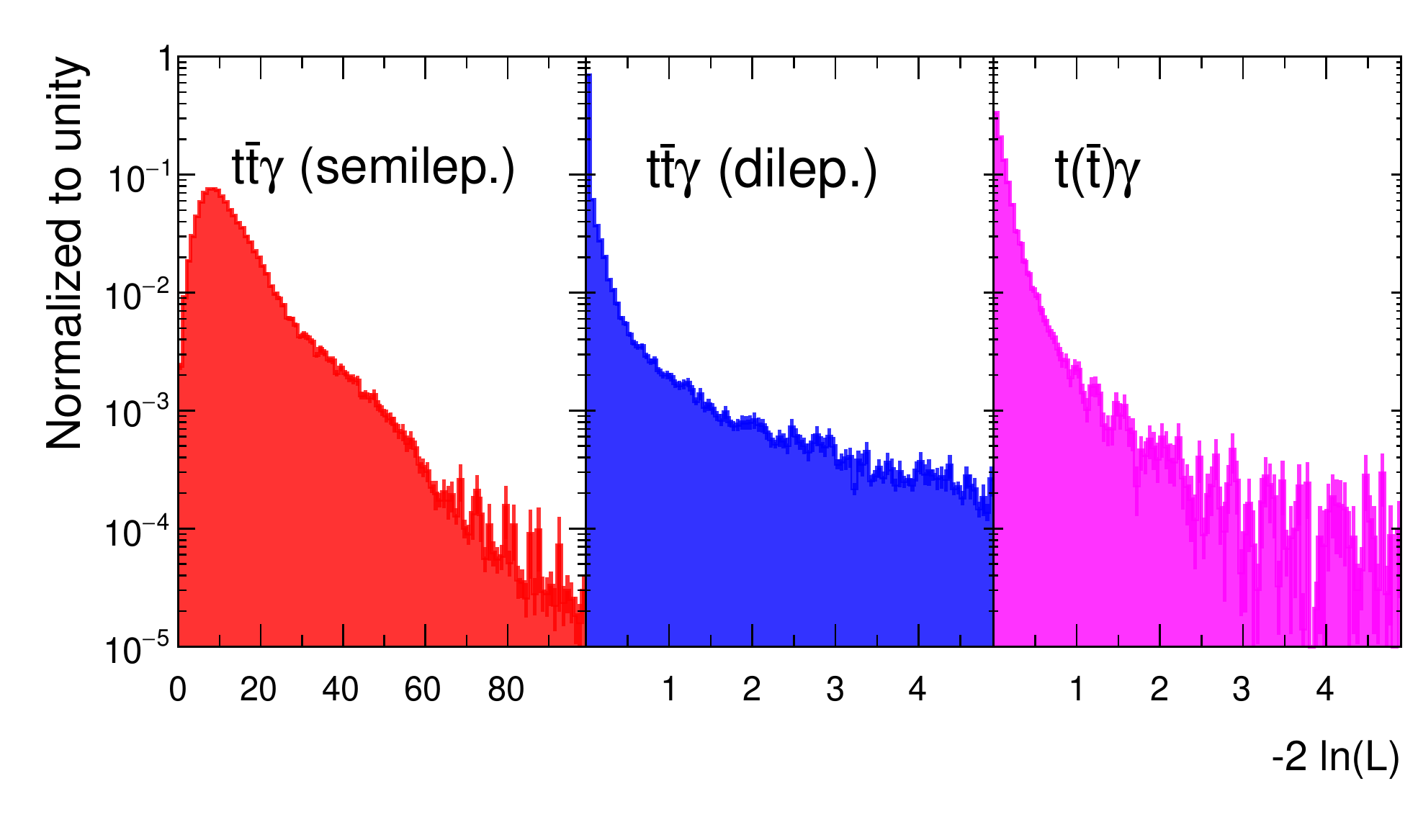}
\caption{\label{fig:disc} Minimum value of the log-likelihood function used in
the kinematic fit in \ttg semileptonic (left), \ttg dilepton
(middle) and \stg (right) events.}
\end{figure}

The likelihood function is evaluated for each
possible assignment of reconstructed leptons and jets to the hard-process
particles. The likelihood minimization is performed in 
each permutation of selected leptons, jets and photons in event. 
The permutations include the photon's origin assignment to
various particles involved in the hard process. 
In order to account for
possible resolution effects in the reconstruction process, a set of the 
so-called transfer functions (TFs) is used that establishes the
correspondence between the generated particles and their reconstructed counterparts.
A TF is defined as the relative difference between the truth-level and
the reconstructed quantities, measured in each considered
category of the reconstructed objects, including jets, \met, leptons and photons, 
as presented in Fig.~\ref{fig:tf}. The use of TFs maximizes the
probability of finding a viable solution for \nupz.
The TFs are used as probability density functions to perform
the generation of pseudo-experiments in each event to define the four-momenta of reconstructed
objects, and in each of these variations the corresponding likelihood
value is computed. Typically, about 100 iterations are needed to find the correct minimum 
in events containing one neutrino. The kinematic fit in dilepton \ttg events uses the 
same minimization algorithm involving TFs with an additional 
minimization step performed in MINUIT~\cite{JamMin}. This is done in order to 
improve the convergence of the minimization process when dealing with
the more complex two-neutrino case of dilepton events. The system of
equations used to derive the solution for \nupz can be resolved in all
relevant event topologies with an efficiency close to 100\%, with some 
dependence on the number of pseudo-experiments.
Permutation that gives the minimum value of the log-likelihood function is
selected to correspond to the best assignment of reconstructed leptons,
jets and photons, to particles generated at parton level. The fitted 
values of the likelihood are shown in
Fig.~\ref{fig:disc}. As described above, the minimum value of the
likelihood is close to zero in $\simeq$~99\% of
selected events in the dilepton case, because the performed fit has no degrees of freedom. 
The corresponding results in \ttg semileptonic events tend to diverge from
zero because the relevant system of equations is overconstrained in the fit.

\section{Results}
\label{sec:results}

\begin{figure}[!htbp]
\centering
\includegraphics[width=1.0\textwidth]{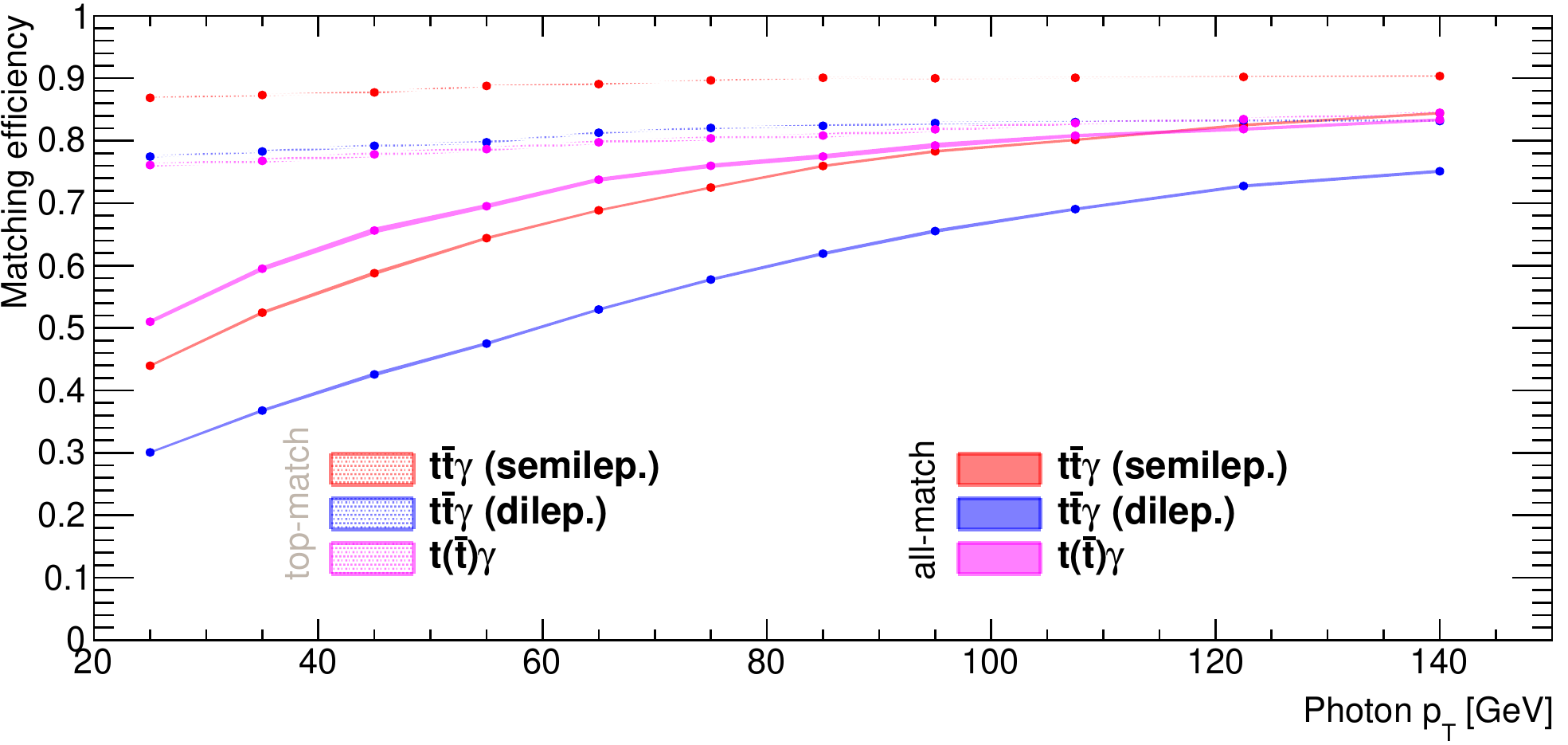}
\hfill
\includegraphics[width=.49\textwidth]{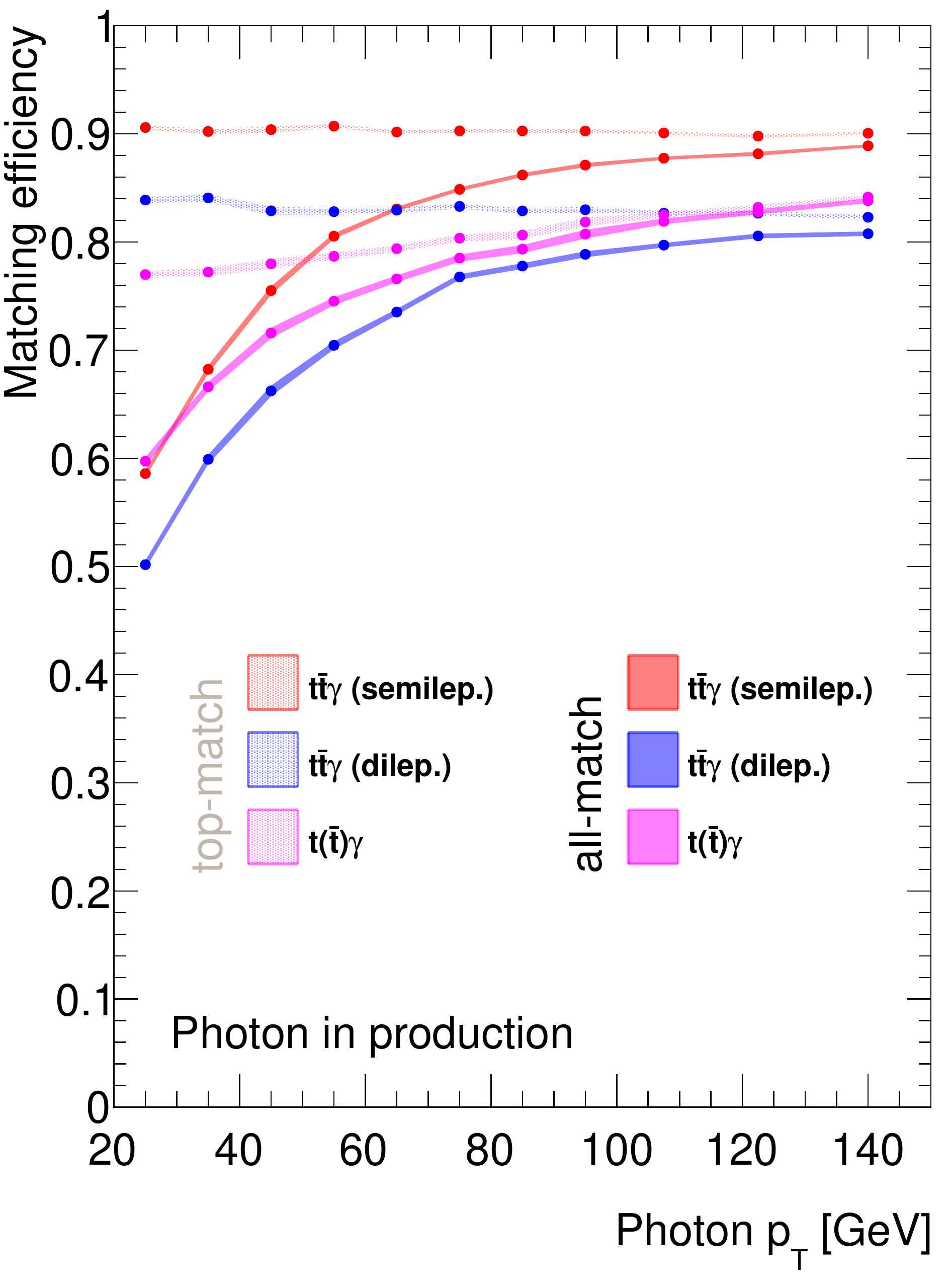}
\hfill
\includegraphics[width=.49\textwidth]{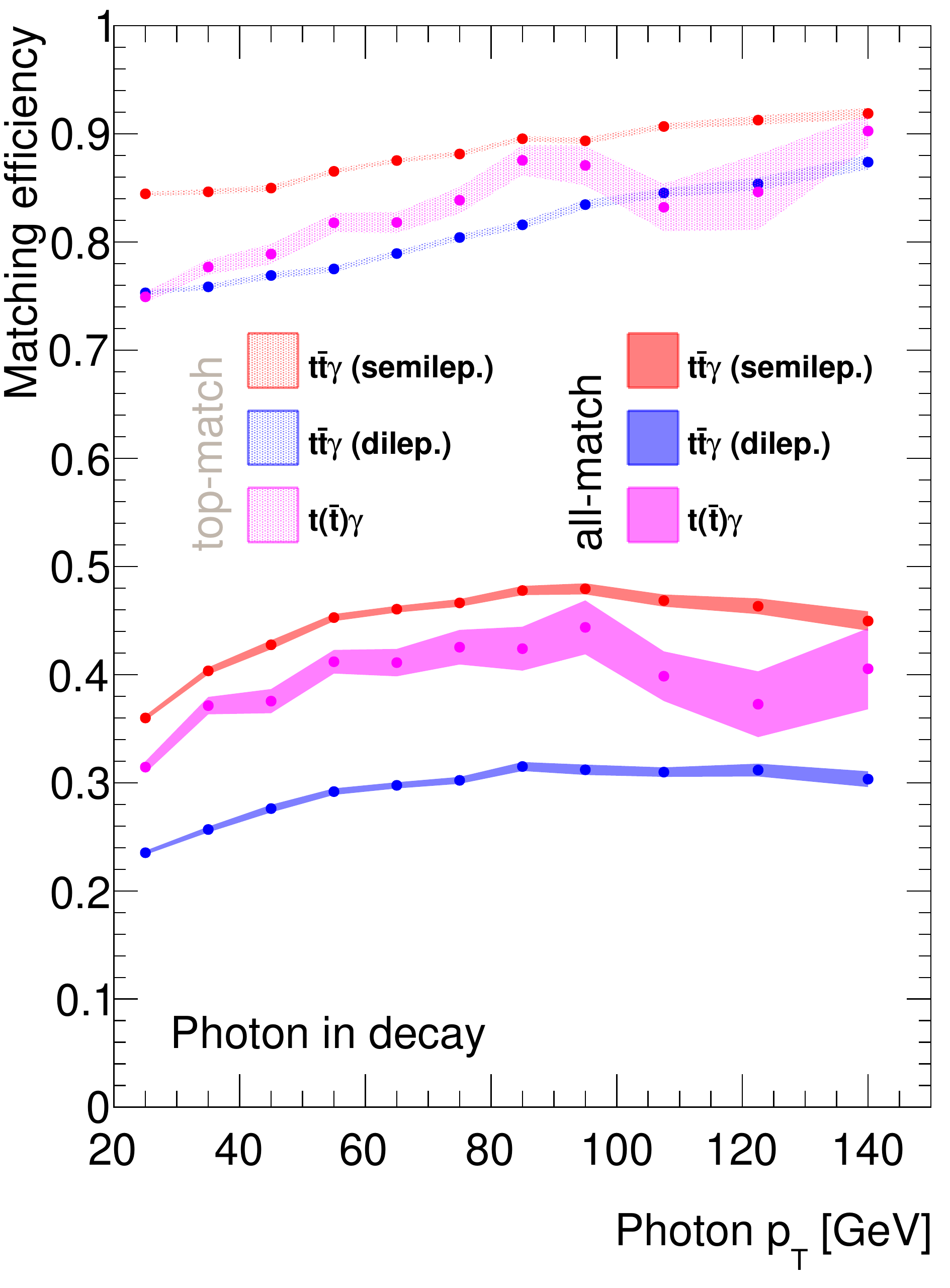}
\caption{\label{fig:eff} Efficiency of correct assignment between
parton- and reco-level objects in top quark events with photon emission (top), 
as well as in events where a photon is radiated
either in the production process (bottom left) or in a decay of a top quark
(bottom right). The efficiencies are shown at central values of bins
defined in the photon \pt and are presented for the correct
matching of the decay products of the produced top quarks
(``top-match'') and when, in addition, the photon is required to be
correctly identified (``all-match''). Colored bands represent the statistical uncertainty.}
\end{figure}

The performance of the proposed method is studied by measuring the
efficiency of the correct assignment of reconstructed objects to
generated leptons, quarks and photons. In order to avoid accounting for any 
inefficiencies due to event reconstruction effects, the selected events 
must have the parton-level particles matched within \dr = 0.4 to the corresponding 
reconstructed objects. The measured efficiencies are presented in
Fig.~\ref{fig:eff} and correspond to the correct matching between the
truth-level generated particles from the decays of top quarks and the
reconstructed objects (``top-match''),
as well as when all generated parton-level final state particles
(leptons, quarks and photons) are correctly matched (``all-match'').

The reconstructed photon in \ttbar events can be assigned to 
either the top quark or a W boson decay, or to any other particle not
involved in the top quark decay process. Since there are two top quarks
and two W bosons produced in \ttbar events, there are four categories in
total corresponding to the photon origin in top quark decay chain. The 
fifth category represents the photon radiation 
from any charged particle involved in the process, excluding the top quark decay products. 
A photon is radiated in ``decay'', if it is assigned to a 
top quark or a W boson decay. It is considered as radiation in
``production'', otherwise.

The kinematic event reconstruction represents the main handle on
the identification of the photon origin in the photon \pt range below
$\simeq$~200 GeV. At higher energies, the photon is predominantly
emitted in production, consistent with the results
obtained from the kinematic fit.

When a photon is radiated in production, the top-match efficiency reaches the
values of $\simeq$~90\% and $\simeq$~83\% in the semileptonic and
dilepton \ttg production channels, respectively, and is rather stable with respect to
the photon \pt. In the case of \stg process, the light-flavour
quark-jet is included in the jet-parton permutations when assigning the b-quark
jet to the corresponding generator b quark. 
There is no kinematic requirement that is used in the fit for the recoiling jet
in the \stg process, which leads to a lower b-quark
matching efficiency. In the \ttg semileptonic events the light-flavour
quark-jets are used in the
calculation of the top quark mass in hadronic W decays, therefore
constraining the allowed parameter phase space of the b-quark jet kinematics.
When additionally requiring the photon to be correctly
assigned to the corresponding hard-process particle, the
overall matching efficiency is found to be $\simeq$~65\% in the region of
the low photon \pt in \ttg semileptonic and \stg events, while it
reaches $\simeq$~50\% in the \ttg dilepton channel. The kinematic
reconstruction in the presence of two neutrinos is associated with
larger systematic effects arising from \met and jet reconstruction,
with respect to the final states involving a single neutrino. Therefore, the
measured matching efficiencies for the reconstructed photon are
generally smaller in the \ttg dilepton channel than in other
considered event topologies. At higher values of the photon \pt, the
matching efficiency reaches a plateau around 100 GeV, because the photon
emission from production becomes dominant, with only a small
contribution from the photon arising from the top quark decay.

In the case of photon radiation from the top quark decay, 
the measured top-match and the all-match efficiencies
increase with the photon \pt. 
The effect of including a radiated photon in the kinematic reconstruction becomes
increasingly pronounced at higher values of the photon \pt, where the 
experimental energy resolution effects in the object reconstruction 
become less important.
The probability to select a permutation that
corresponds to the proper matching of all hard-process particles is
$\simeq$~45\% in \ttg semileptonic events, while it reaches
$\simeq$~30\% and $\simeq$~40\% in \ttg dilepton and \stg events,
respectively. In case of the hadronic decay of a W boson, the reconstructed 
jets are allowed to be matched to either of the two
quarks from the W boson decay, but not to the same one.

\begin{figure}[!htbp]
\centering
\includegraphics[width=.49\textwidth]{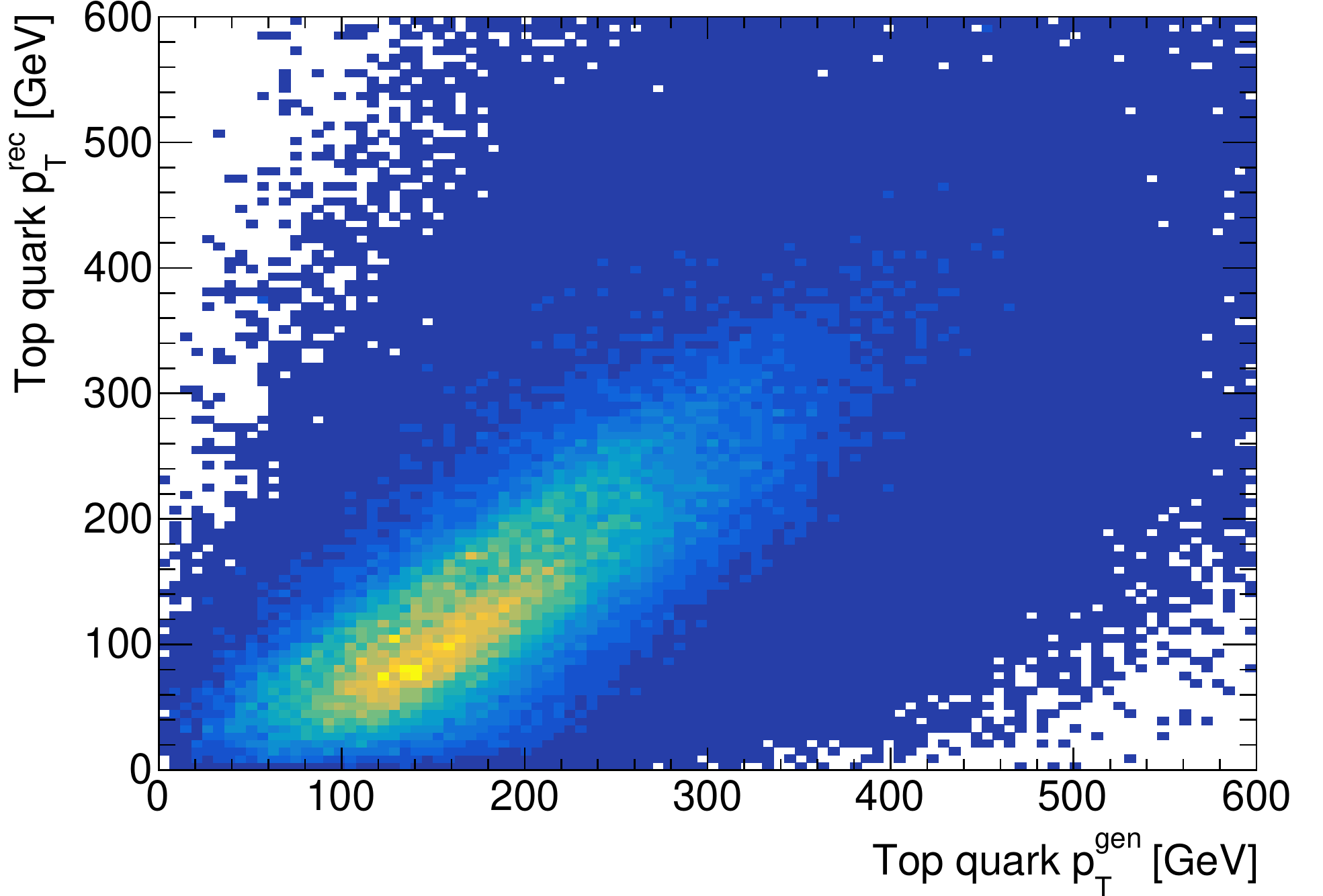}
\hfill
\includegraphics[width=.49\textwidth]{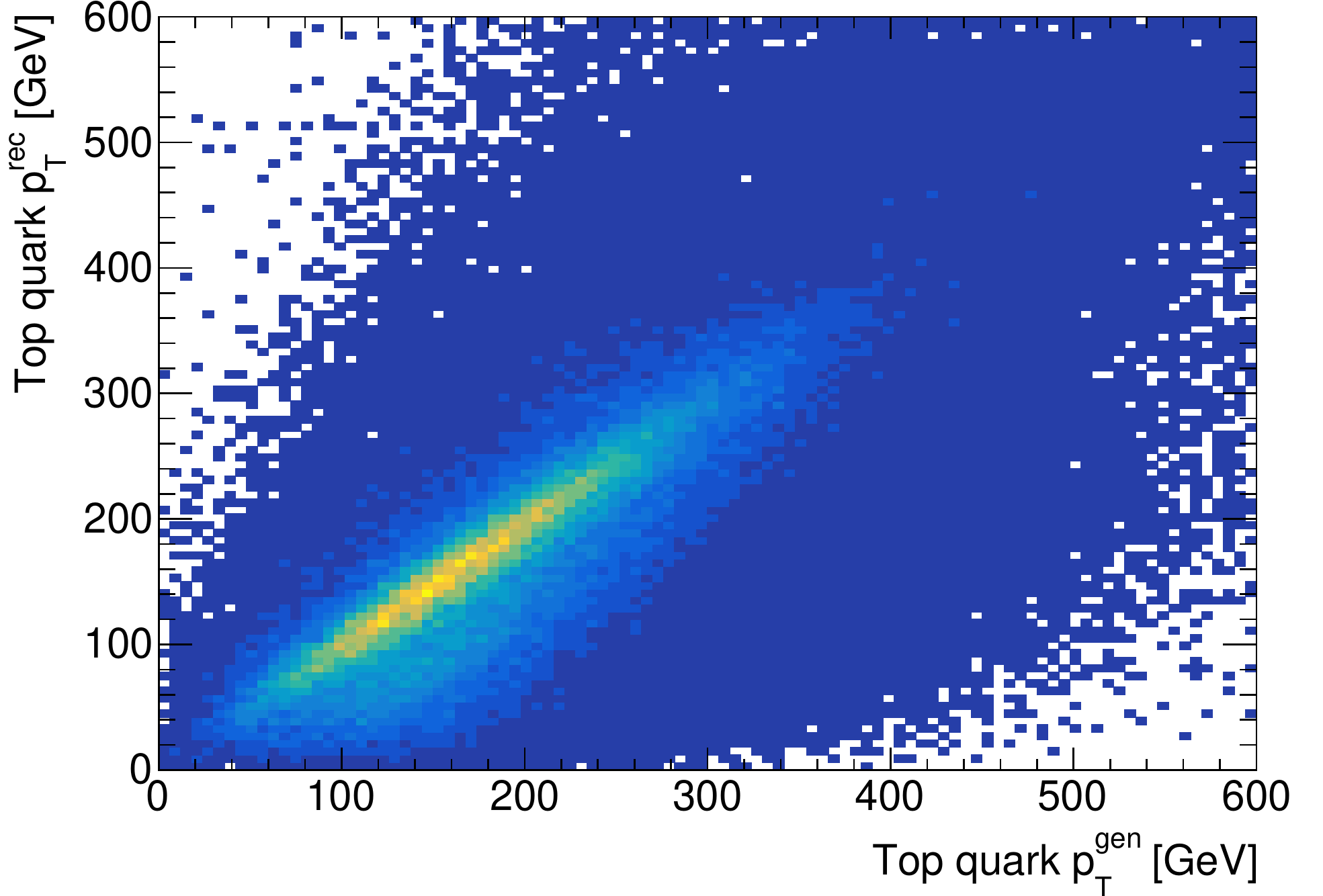}
\caption{\label{fig:ttg_top1Pt} Transverse momentum of the
top quark that corresponds to a W boson leptonic decay as obtained in
semileptonic \ttg events with a radiated photon of \pt > 50 GeV. 
A correlation pattern is shown between the
generated and the reconstructed values, when excluding (left) or including (right) 
the reconstructed photon in the kinematic fit. 
}
\end{figure}

\begin{figure}[!htbp]
\centering
\includegraphics[width=.49\textwidth]{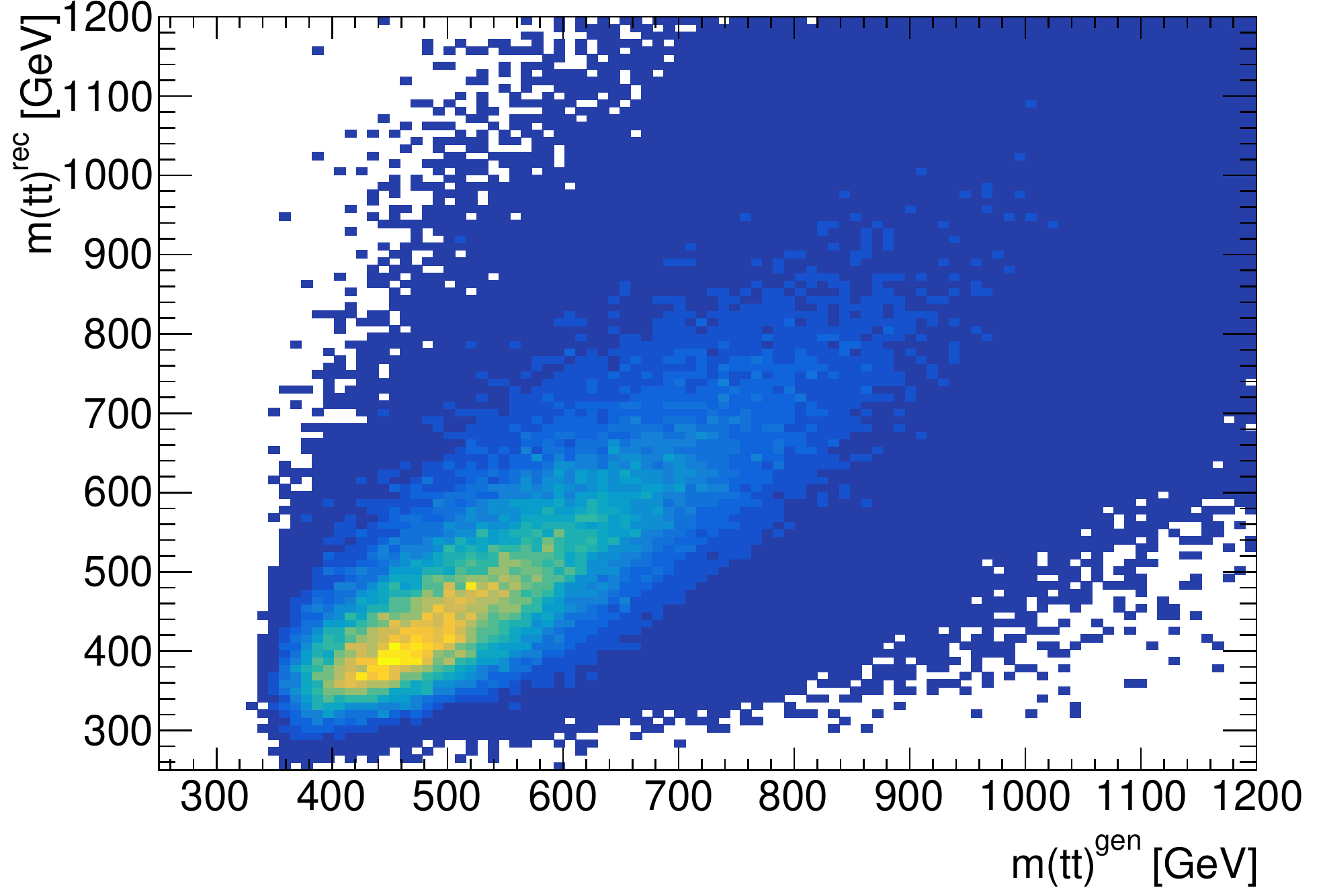}
\hfill
\includegraphics[width=.49\textwidth]{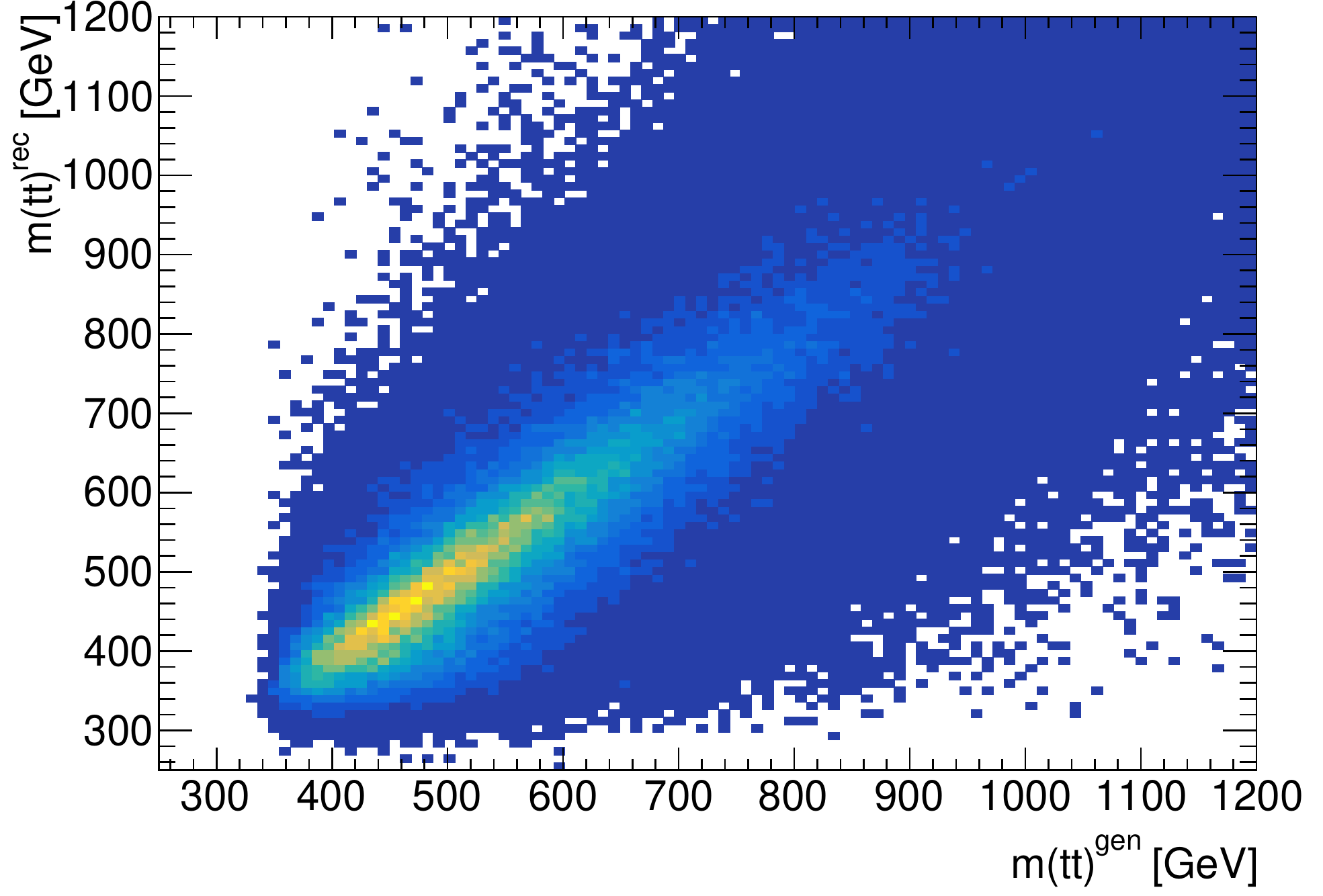}
\caption{\label{fig:ttg_top2Pt} Invariant mass of two top quarks in semileptonic \ttg events
with a radiated photon of \pt > 50 GeV.
A correlation pattern is shown between the
generated and the reconstructed values, when excluding (left) or including (right)
the reconstructed photon in the kinematic fit. 
}
\end{figure}

\begin{figure}[!htbp]
\centering
\includegraphics[width=.49\textwidth]{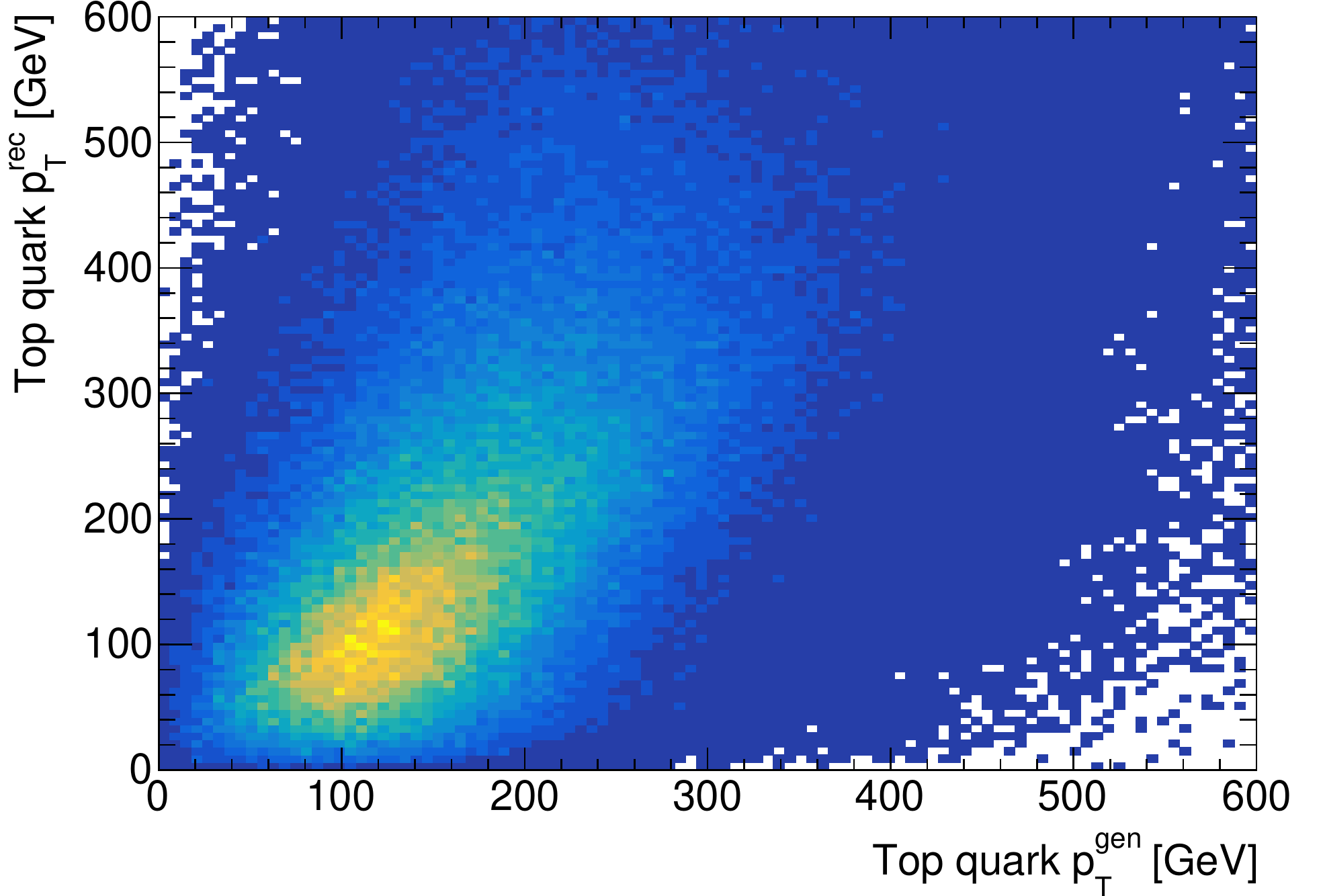}
\hfill
\includegraphics[width=.49\textwidth]{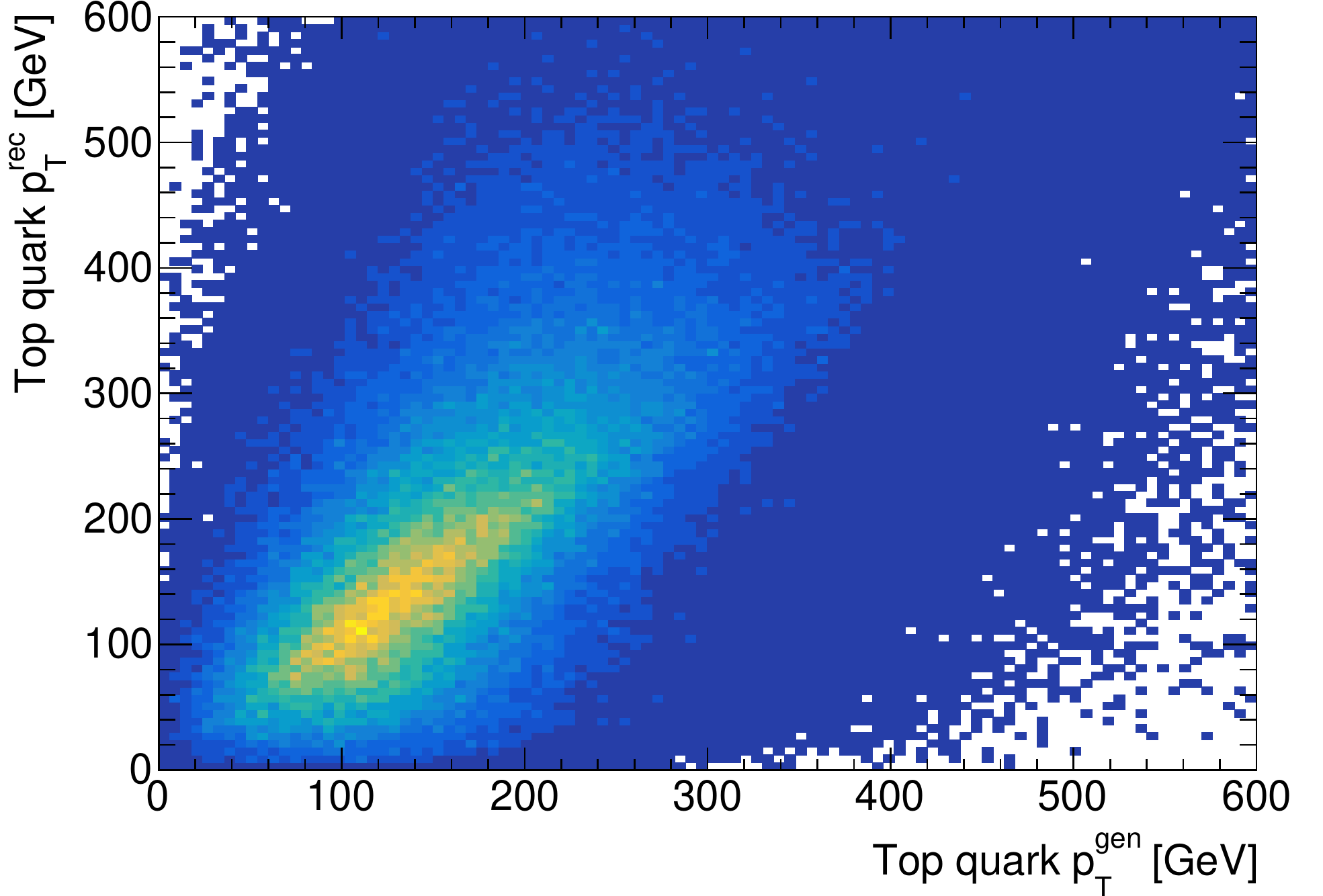}
\caption{\label{fig:ttg_dilepton_topPt} Transverse momentum of top quarks in dilepton 
\ttg events with a radiated photon of \pt > 50 GeV.
A correlation pattern is shown between the
generated and the reconstructed values, when excluding (left) or including (right)
the reconstructed photon in the kinematic fit. 
}
\end{figure}

\begin{figure}[!htbp]
\centering
\includegraphics[width=.49\textwidth]{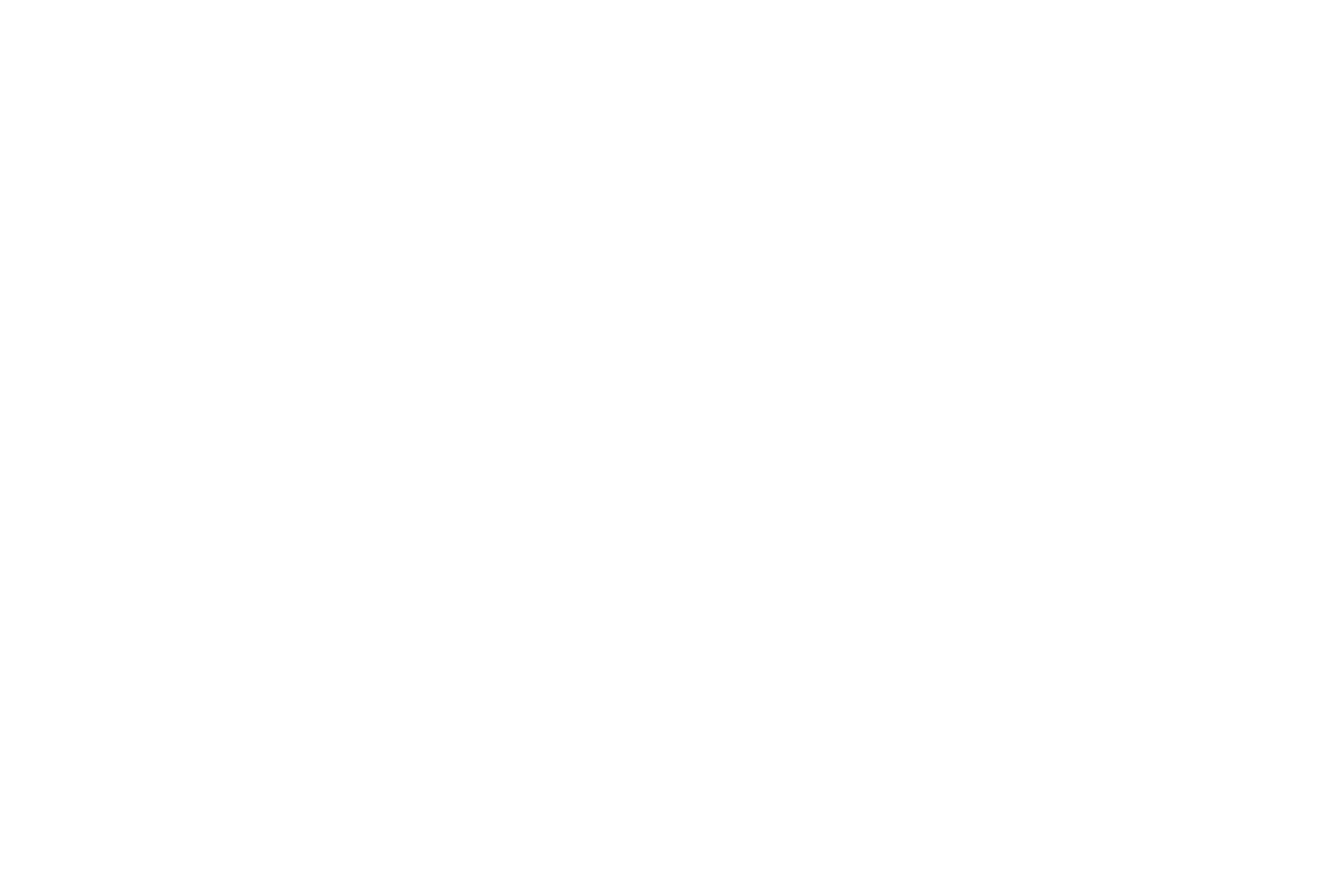}
\hfill
\includegraphics[width=.49\textwidth]{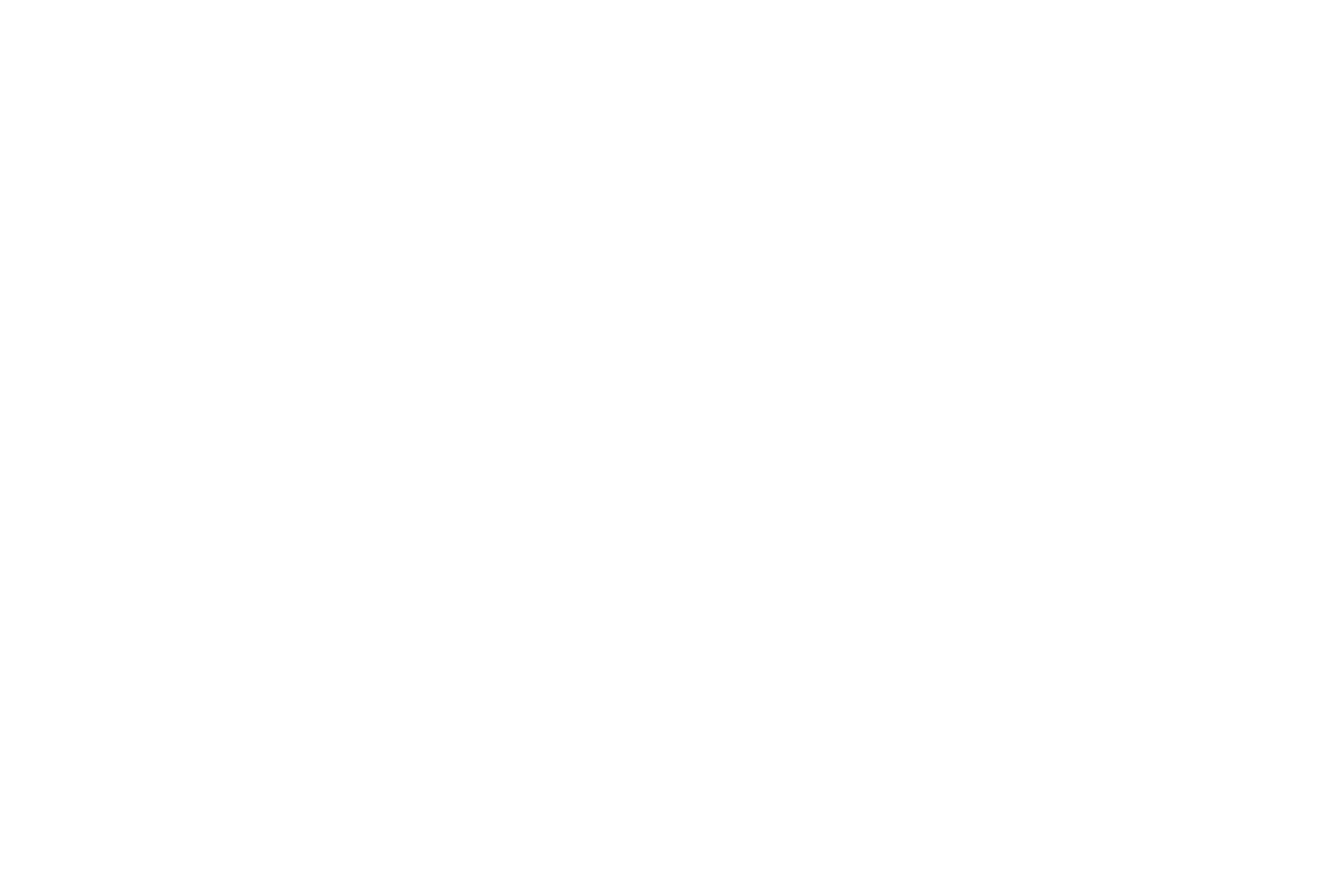}
\caption{\label{fig:tg_top1Pt} Transverse momentum of top quarks
in \stg events with a radiated photon of \pt > 20 GeV.
A correlation pattern is shown between the
generated and the reconstructed values, when excluding (left) or including (right) 
the reconstructed photon in the kinematic fit. 
}
\end{figure}

\begin{figure}[!htbp]
\centering
\includegraphics[width=.49\textwidth]{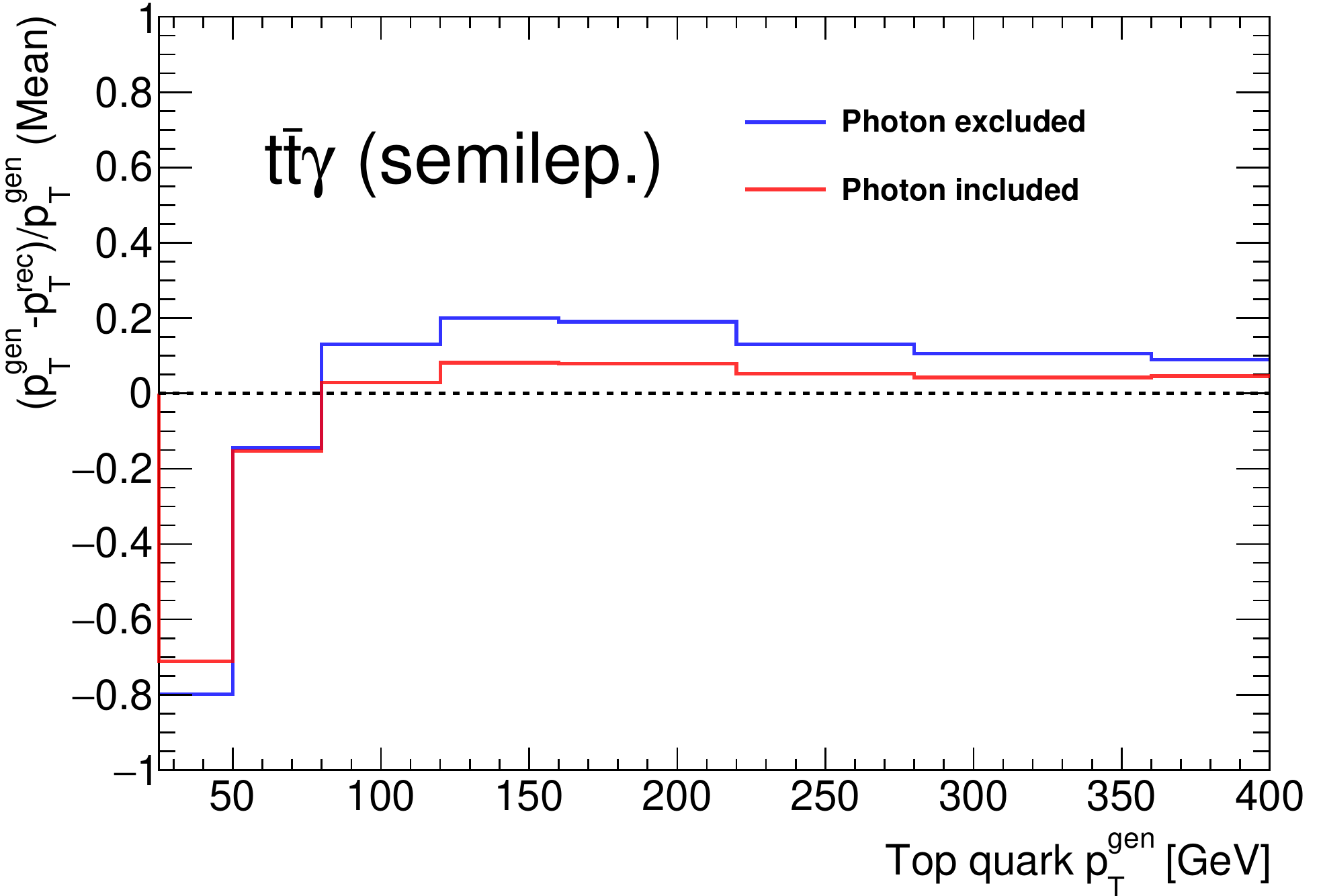}\hfill
\includegraphics[width=.49\textwidth]{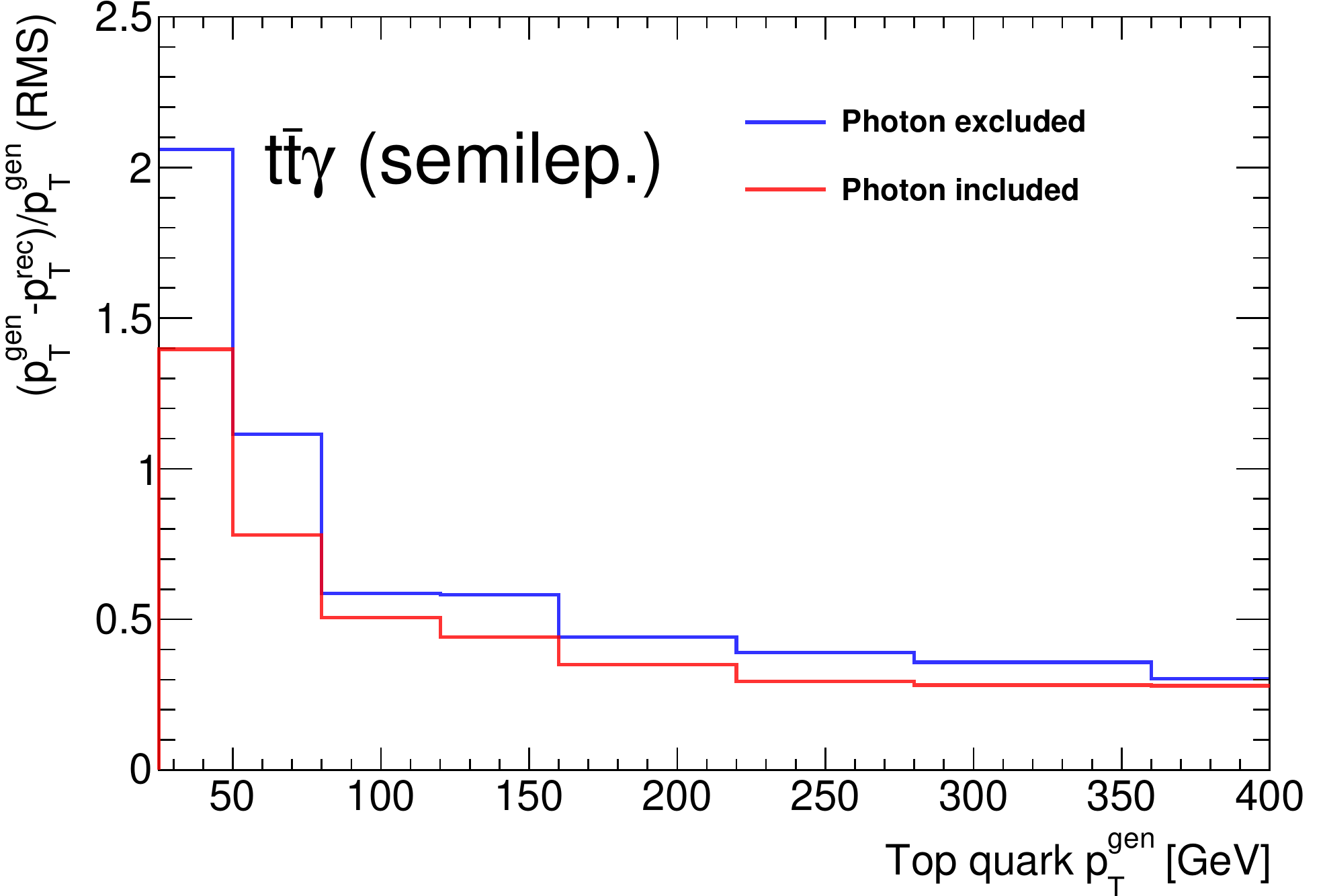}
\includegraphics[width=.49\textwidth]{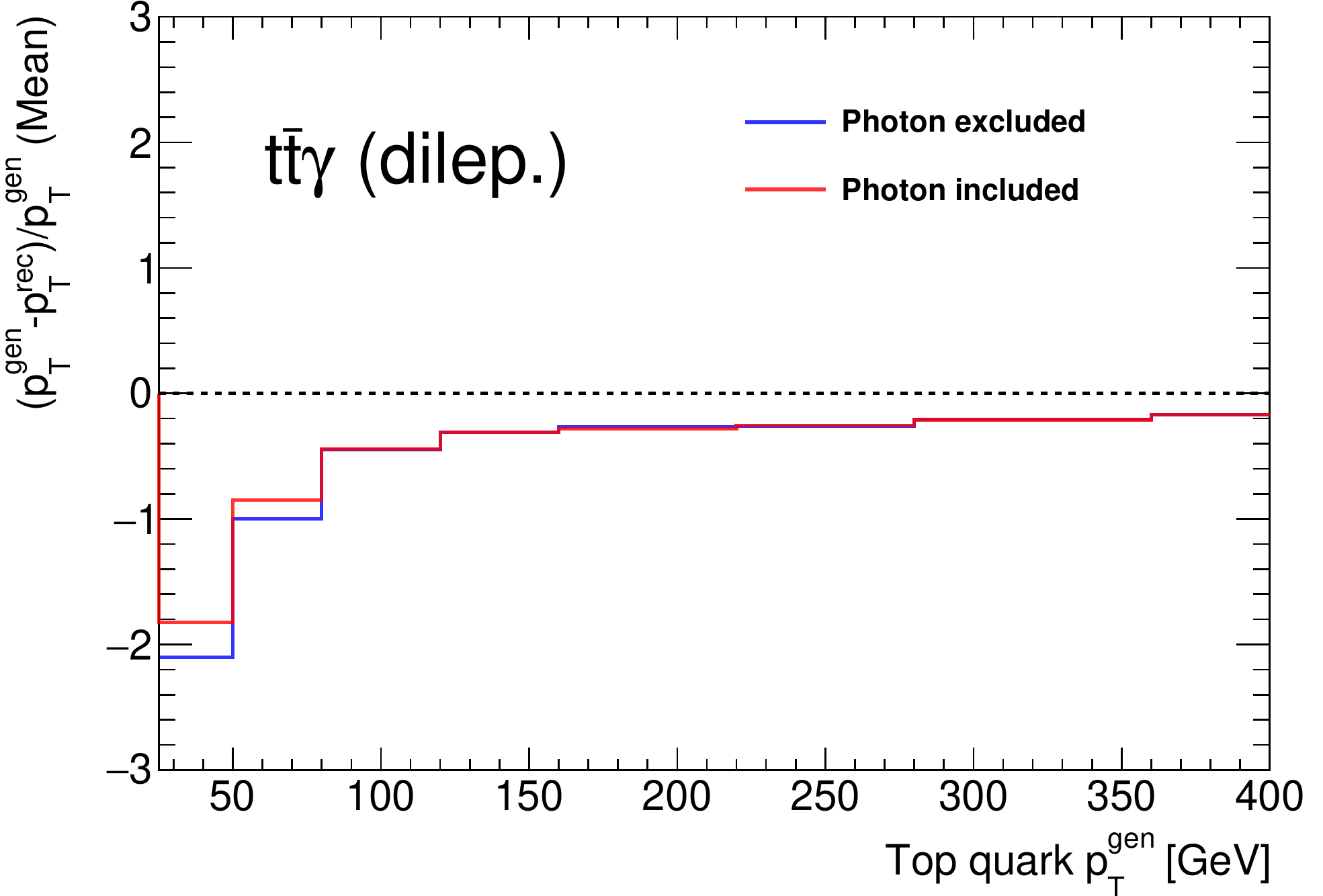}\hfill
\includegraphics[width=.49\textwidth]{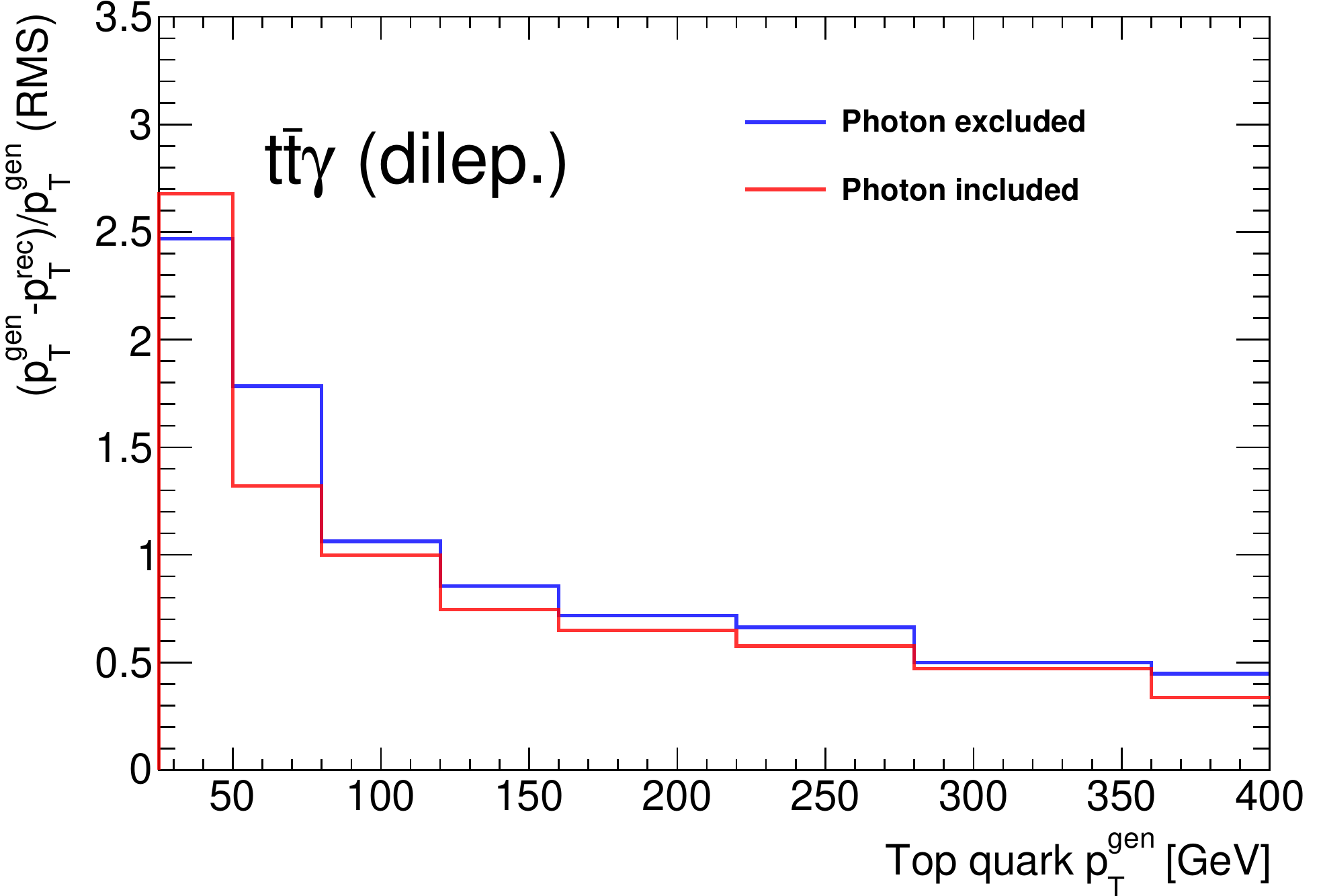}
\includegraphics[width=.49\textwidth]{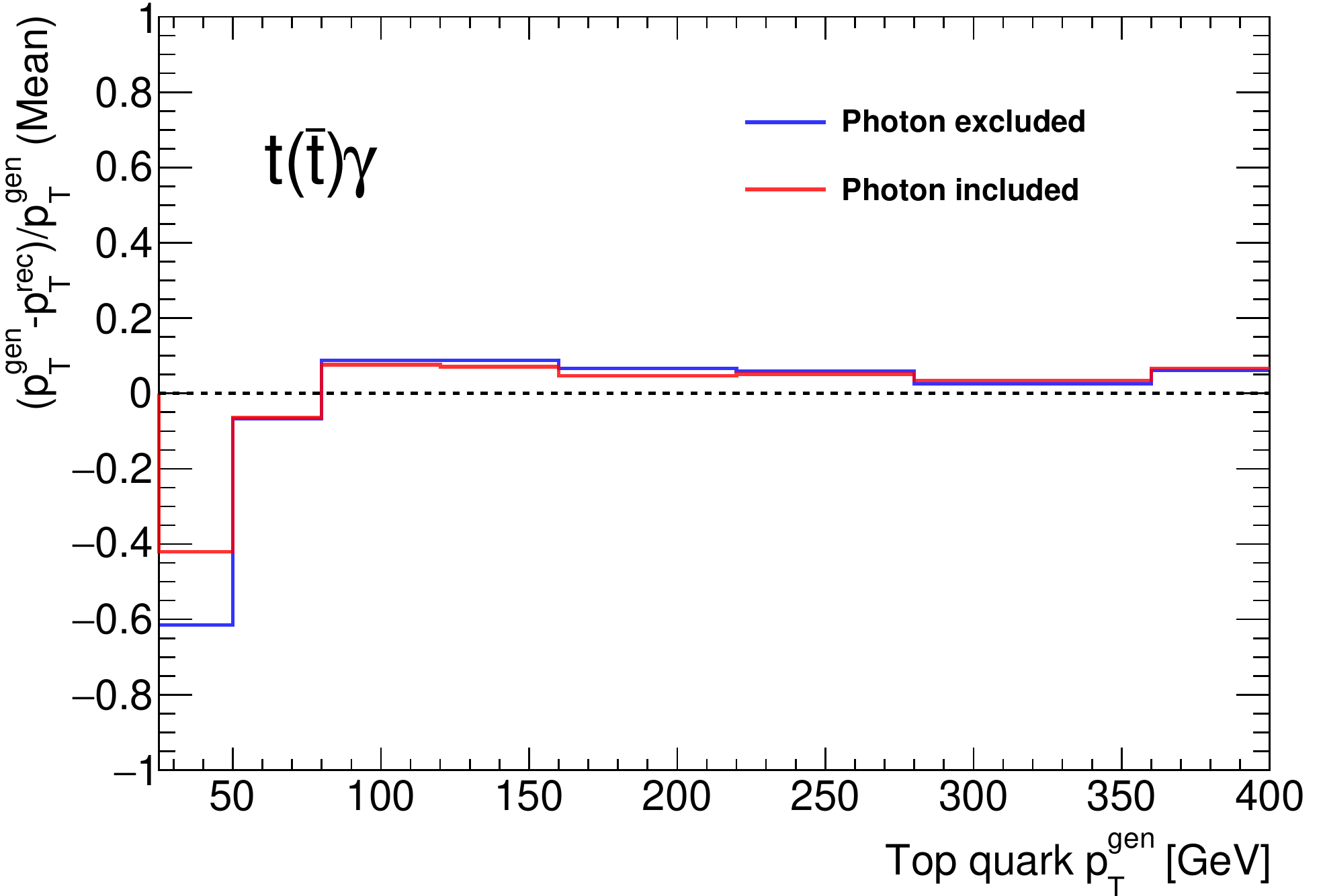}\hfill
\includegraphics[width=.49\textwidth]{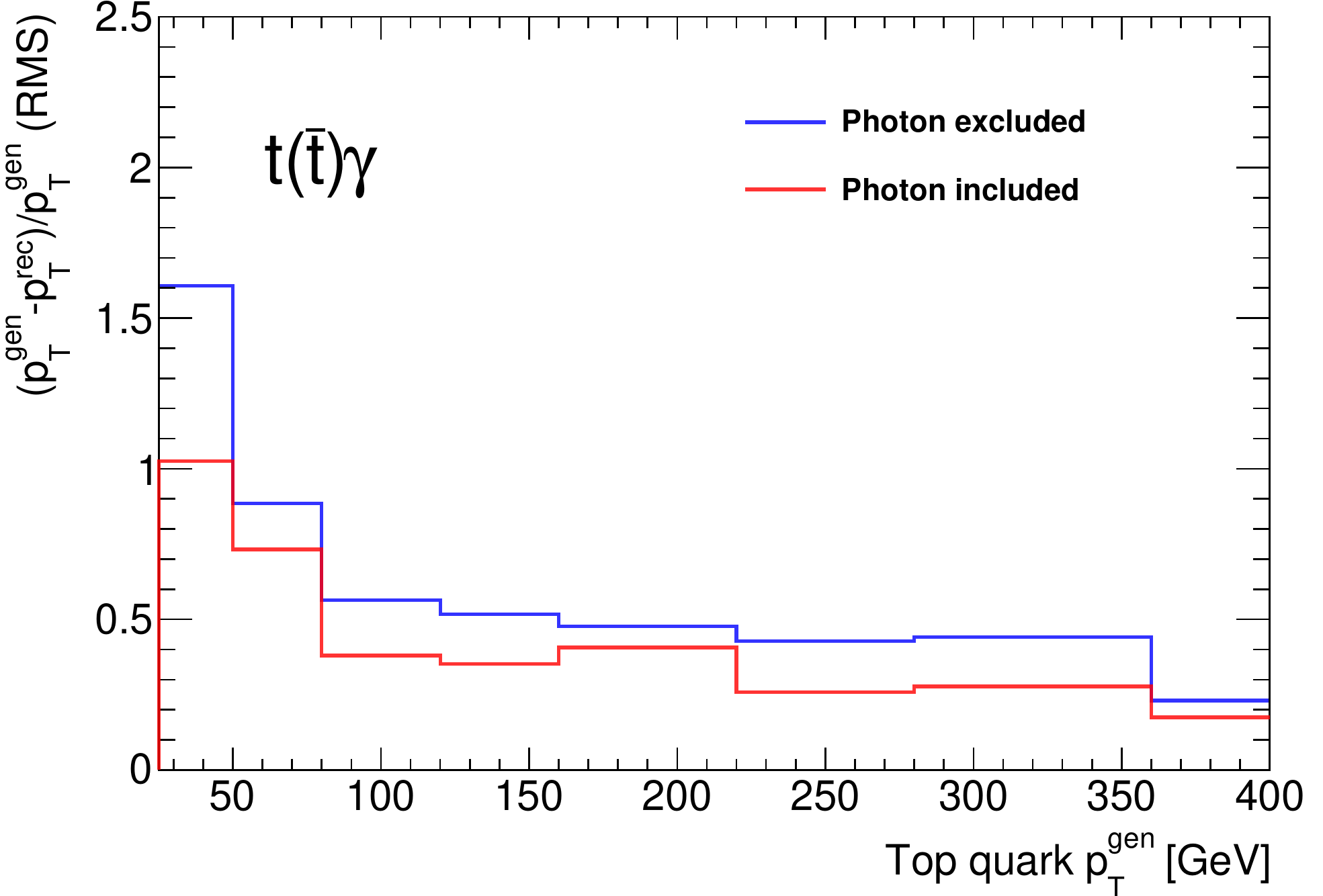}
\caption{\label{fig:res} The measured mean values and the corresponding 
variances for the relative difference in the transverse
momentum of reconstructed and generated top quarks in 
semileptonic \ttg, dilepton \ttg and \stg events. The comparison is
presented for two scenarios: when a radiated photon
of \pt > 50 (20) GeV in \ttg (\stg) events is either
excluded or included in the kinematic fit.
}
\end{figure}

A correct assignment of reconstructed objects to the hard-process
particles allows to define kinematic properties of produced 
top quarks. The comparison between the reconstructed and the generated top
quark \pt in \ttg semileptonic events is presented in
Fig.~\ref{fig:ttg_top1Pt}. A similar comparison for an hadronically
decaying top quark is shown in Fig.~\ref{fig:ttg_top2Pt}. The results
from the application of kinematic reconstruction to the \ttg dilepton
and \stg events are presented in Figs.~\ref{fig:ttg_dilepton_topPt}
and~\ref{fig:tg_top1Pt}, respectively.
The comparisons are shown for the case, where the photon is omitted in the kinematic
event reconstruction, as well as for the case where it is included in the fit. 
When the photon is not accounted for in the reconstruction, the correlation between the
reconstructed and generated top quark \pt appears to be clearly
biased and correlated to the minimum photon \pt requirement that is
applied in selected events, while the inclusion of the photon leads to a significantly improved
resolution in derived kinematic variables.
The results are also presented as the mean values and the corresponding
variances in the measured top quark \pt in Fig.~\ref{fig:res}.

\section{Summary}
\label{sec:summary}

A method for the reconstruction of the top quark event kinematics in the presence of
photon radiation is presented. The kinematic reconstruction is applied to events 
with the production of top quark pairs in association with a photon, and to events
with an associated production of single top quarks with a photon. 
The inclusion of a radiated photon in a kinematic fit allows to
significantly improve the precision of reconstructed
top quark kinematic observables, when compared to the event generator
truth-level information. The proposed method can be used in experimental studies
of the processes with top quarks with an additional radiation of photons.
It aims at improving the precision of the differential cross section
measurements, especially in the low photon \pt region, as well as
at providing an additional handle for the top quark charge asymmetry
measurement. The presented algorithm of kinematic reconstruction only
relies on the knowledge of the top quark and the W boson masses and
therefore can be applied to other processes with similar decay
topologies where photon radiation effects are of importance.

\acknowledgments

We would like to warmly thank Dennis Schwarz, Francisco Yumiceva,
Nadjieh Jafari and Daniel Noonan for carefully reading the manuscript and for providing
an invaluable set of comments on the presented study. We also thank
Robert Sch\"{o}fbeck and Lukas Lechner for numerous
discussions on the subject of an experimental study of the process 
with the production of top quark pairs in association with photons
over the past years.

\end{document}